# Generating Loop Invariants by Computing Vanishing Ideals of Sample Points[⋆]

Bin Wu [a,c]  Liyong Shen [b]  Min Wu [a]  Zhengfeng Yang [a]
Zhenbing Zeng [a]

[a] *Shanghai Key Laboratory of Trustworthy Computing,
East China Normal University, Shanghai 200062, China*

[b] *School of Mathematical Sciences, Graduate University of CAS,
Beijing 100049, China*

[c] *Shanghai University of Finance and Economics Zhejiang College,
Jinhua 321019, China*

**Abstract**

Loop invariants play a very important role in proving correctness of programs. In this paper, we address the problem of generating invariants of polynomial loop programs. We present a new approach, for generating polynomial equation invariants of polynomial loop programs through computing vanishing ideals of sample points. We apply rational function interpolation, based on early termination technique, to generate invariants of loop programs with symbolic initial values. Our approach avoids first-order quantifier elimination and cylindrical algebraic decomposition(CAD). An algorithm for generating polynomial invariants is proposed and some examples are given to illustrate the algorithm. Furthermore, we demonstrate on a set of loop programs with symbolic initial values that our algorithm can yield polynomial invariants with degrees high up to 15.

## 1. Introduction

Loop invariant generation plays a central role in program verification. An *invariant* of a loop program at a location is an assertion over the program variables that is true of

[⋆] This research was partly supported by the NSFC projects No. 90718041, 10801052, 10901055, and the Shanghai Natural Science Foundation under Grant (09ZR1408800).

*Email addresses:* binwu.cs@gmail.com (Bin Wu), lyshen@gucas.ac.cn (Liyong Shen), mwu@sei.ecnu.edu.cn (Min Wu), zfyang@sei.ecnu.edu.cn (Zhengfeng Yang), zbzeng@sei.ecnu.edu.cn (Zhenbing Zeng).

any program state reaching the location. Loop invariants are helpful for program analysis and verification.

Since the late seventies of the 20th century, many methods have been proposed to generate loop invariants. In (German and Wegbreit, 1975; Katz and Manna, 1976; Wegbreit, 1974, 1975), difference equation solving techniques were used to generate loop invariants. However, this technique is difficult to apply in general, since difference equations are generally hard to solve. In (Karr, 1976; Cousot and Cousot, 1977; Cousot and Halbwachs, 1978), abstract interpretation techniques were applied to finding linear equation or inequality invariants.

Based on some previous work, Müller-Olm and Seidl (Müller-Olm and Seidl, 2004a,b) generated polynomial equation invariants of loop programs with affine assignments using linear algebra techniques.

Recently, the constraint-based methods become dominant in invariant generation. These methods require to preset a template of the invariant as a polynomial equation or inequality with unknown coefficients, and the initiation and consecution conditions for the invariants generate constraints on the unknown coefficients. Then a solution to the constraint system yields invariants. In (Colón et al., 2003), Farkas' Lemma was applied to generating linear inequality invariants using non-linear constraint solving. In (Kapur, 2004), Kapur proposed an approach based on quantifier elimination to generate polynomial equation invariants. In (Sankaranarayanan et al., 2004), the polynomial-scale consecution of inductive invariants was first defined, and the polynomial equation invariants satisfying polynomial-scale consecution were computed using an extended Gröbner basis algorithm. (Rebiha et al., 2008) proposed a complete method using multi-parametric constraints to generate polynomial invariants that satisfy polynomial-scale consecution. To generate polynomial equation or inequality invariants of loop programs with guard conditions and branches, Chen et al. (Chen et al., 2007) applied the techniques of solving semi-algebraic systems.

Rodríguez-Carbonell and Kapur (Rodríguez-Carbonell and Kapur, 2004, 2007) first proved that the polynomial equation invariants have the algebraic structure of an ideal, and they proposed a fixpoint procedure for finding all polynomial equation invariants using Gröbner bases and quantifier elimination. (Kovács, 2007; Kauers and Zimmermann, 2008; Kovács, 2008) proposed complete algorithms to generate polynomial equation invariants for a restricted class of linear (P-solvable) loops.

In this paper, by computing vanishing ideals of program sample points, we present a new method for generating polynomial invariants of polynomial loop programs in which the guard conditions and assignments are polynomials in the program variables.

Recall that a multivariate polynomial in $n$ variables with total degree bound $e$ has at most $\binom{n+e}{n}$ distinct terms. Therefore, to compute the invariants with a given degree bound $e$, we first get no more than $\binom{n+e}{n}$ sample points by executing the loop program, where $n$ is the number of program variables. Then we apply Buchberger-Möller algorithm to compute the vanishing ideal of these sample points as candidate invariants (a candidate may not be real invariant). Subsequently, the problem of verifying the candidate invariants can be translated into that of determining divisibility between multivariate polynomials, and a practical probabilistic method is presented to exclude non-invariants quickly. Finally, we can either generate the polynomial invariants or conclude that the polynomial invariants with degree $\leq e'$ do not exist, where $e'(\leq e)$ is the minimal degree of the polynomials in the vanishing ideal. Moreover, rational function interpolation method, combining variable by variable interpolation with early termination technique (Kaltofen and Yang, 2007) is applied to generating invariants of loop programs with symbolic initial values.

The rest of the paper is organized as follows. In Section 2, we recall the notions of vanishing ideals for finitely many points, transition systems and (inductive) invariants. In Section 3, we present an efficient method to generate polynomial equation invariants for polynomial loop programs with initial values, and in Section 4, an algorithm and some examples are given. We conclude our results in Section 5.

## 2. Notation and Definitions

*2.1. Vanishing Ideals of Finitely Many Points*

This section contains a collection of definitions and facts about vanishing ideals of finitely many points.

Throughout this paper, let $\mathcal{K}$ be a (commutative) field of characteristic zero, $\mathcal{K}[x_1,\ldots,x_n]$ be the ring of polynomials in $n$ indeterminates $x_1,\ldots,x_n$ over $\mathcal{K}$, the term order $\sigma$ in $\mathcal{K}[x_1,\ldots,x_n]$ be the graded lexicographic order, and $\deg(f)$ denote the total degree of a polynomial $f \in \mathcal{K}[x_1,\ldots,x_n]$.

**Definition 1** (Ideal of Polynomials). A set $\mathcal{I} \subseteq \mathcal{K}[x_1,\ldots,x_n]$ is an *ideal* in $\mathcal{K}[x_1,\ldots,x_n]$ if for any $f, g \in \mathcal{I}$ and $h \in \mathcal{K}[x_1,\ldots,x_n]$, we have $f + g \in \mathcal{I}$ and $f \cdot h \in \mathcal{I}$.

For $h_1,\ldots,h_r \in \mathcal{K}[x_1,\ldots,x_n]$, we denote by $\langle h_1,\ldots,h_r \rangle$ the smallest ideal containing $h_1,\ldots,h_r$, i.e.

$$\langle h_1,\ldots,h_r \rangle = \left\{ \sum_{i=1}^{r} f_i h_i \mid f_i \in \mathcal{K}[x_1,\ldots,x_n] \right\}.$$

If $\mathcal{I} = \langle h_1,\ldots,h_r \rangle$, we say that $\mathcal{I}$ is an ideal generated by $h_1,\ldots,h_r$ and that $h_1,\ldots,h_r$ is a basis of $\mathcal{I}$.

By Hilbert' Basis Theorem, any ideal in $\mathcal{K}[x_1,\ldots,x_n]$ has a *finite basis*. The classical Buchberger's algorithm (Buchberger, 1985) can be applied to computing Gröbner bases of ideals.

**Definition 2** (Vanishing Ideal of Finitely Many Points). Let $A$ be a finite subset of $\mathcal{K}^n$. The *vanishing ideal* of the point set $A$ is the ideal

$$\mathcal{I}(A) = \{f \in \mathcal{K}[x_1,\ldots,x_n] \mid f(a) = 0, \text{ for all } a \in A\}$$

of all the polynomials that vanish on each point in $A$.

Buchberger and Möller presented an algorithm (Möller and Buchberger, 1982) to compute the reduced Gröbner bases of the vanishing ideals for finitely many points, via Gaussian elimination on a generalized Vandermonde matrix.

**Remark 1.** As stated in (Marinari et al., 1993), Buchberger-Möller algorithm is of polynomial time complexity $O(n^2 s^4)$, where $n$ is the dimension of the affine space $\mathcal{K}^n$ and $s$ is the number of points.

## 2.2. Transition Systems and Invariants

The classical way to represent programs is the use of transition systems.

**Definition 3** (Transition System). A *transition system* T is a tuple $\langle V, L, \mathcal{T}, l_0, \Theta \rangle$, where
- $V = \{x_1, \ldots, x_n\}$ is a finite set of program state variables;
- $L$ is a set of locations;
- $\mathcal{T}$ is a set of transitions, where each transition $\tau \in \mathcal{T}$ is a tuple of the form

$$\langle l_1, l_2, \rho_\tau, g_\tau \rangle,$$

such that
  - $l_1, l_2 \in L$ are the pre- and post- locations of $\tau$, respectively;
  - $\rho_\tau$ is a *transition relation*, i.e., a first-order formula, over $V \cup V'$, where the prime version $V' := \{x'_1, \ldots, x'_n\}$ of $V$ represents the next-state variables. Here $x'_i$ is the new variable introduced to stand for the value of $x_i$ after the assignment. For example, the assignment $x := x + 1$ can be written as $x' = x + 1$;
  - $g_\tau$ is the guard condition of the transition $\tau$ or of $\rho_\tau$. Only if $g_\tau$ holds, the transition can take place.
- Location $l_0 \in L$ is an initial location, and the initial condition $\Theta$ is a first-order formula over $V$.

As an example, a loop program

$$\textbf{where } \Theta$$
$$l: \textbf{ while } guard \textbf{ do}$$
$$x_1 := P_1(x_1, \ldots, x_n);$$
$$\vdots$$
$$x_n := P_n(x_1, \ldots, x_n);$$
$$\textbf{end while} \tag{1}$$

can be translated easily into a transition system

$$\langle \{x_1, \ldots, x_n\}, \{l\}, \{\tau\}, \{l\}, \Theta \rangle$$

where

$$\tau = \langle l, l, x'_1 = P_1(x_1, \ldots, x_n) \wedge \cdots \wedge x'_n = P_n(x_1, \ldots, x_n), guard \rangle. \tag{2}$$

A *polynomial loop* program is a loop program (1) where all the $P_i$ are polynomials in $x_1, \ldots, x_n$, and the guard is represented by a conjunction of polynomial inequalities.

By an *assertion*, we mean a first-order formula over the program variables. A *state* of a transition system is an interpretation of the program variables as values from the corresponding domains. We use the notation $s \models \varphi$ to denote that a state $s$ satisfies an assertion $\varphi$. We will also write $\varphi_1 \models \varphi_2$ for two assertions $\varphi_1, \varphi_2$ to represent that $\varphi_2$ is true at least in all the states in which $\varphi_1$ is true.

Next, we introduce the notions of (inductive) invariants for transition systems.

**Definition 4** (Invariant). Let $T = \langle V, L, \mathcal{T}, l_0, \Theta \rangle$ be a transition system. An *invariant* of the system T at location $l \in L$ is an assertion over $V$ which holds at all reachable states at location $l$. An *invariant* of the system T is an assertion over $V$ that holds at all locations.

**Definition 5** (Inductive Invariant). Let $T = \langle V, L, \mathcal{T}, l_0, \Theta \rangle$ be a transition system and $D$ the domain of assertions. An *assertion map* for T is a map $\eta : L \to D$ that associates each location of the transition system with an assertion. We say that $\eta$ is *inductive* if the *Initiation* and *Consecution* conditions hold:
**Initiation:** $\Theta \models \eta(l_0)$;
**Consecution:** For each transition $\tau = \langle l_1, l_2, \rho_\tau, g_\tau \rangle$, we have

$$\eta(l_1)(V) \wedge \rho_\tau(V, V') \wedge g_\tau(V) \models \eta(l_2)(V'), \tag{3}$$

where $\eta(l_2)(V')$ represents the assertion $\eta(l_2)$ with the current state variables $x_1, \ldots, x_n$ replaced by the next state variables $x'_1, \ldots, x'_n$, respectively.

It is a well-known result (Floyd, 1967) that if $\eta$ is an inductive assertion map then $\eta(l)$ is an invariant at location $l$ for each $l \in L$.

**Remark 2.** When no confusion arises, we write the condition (3) simply as

$$\eta(V) \wedge \rho_\tau(V, V') \wedge g_\tau(V) \models \eta(V').$$

In this paper, we are interested in finding inductive invariants of the form $p(V) = 0$, where $p(V)$ is a polynomial in the program variables. For brevity, we shall use $\eta(V)$ to denote both the assertion $p(V) = 0$ and the polynomial $p(V)$.

In the sequel, we will use the following stronger but more practical consecution condition defined in (Sankaranarayanan et al., 2004, Definition 13) or (Rebiha et al., 2008, Definition 4).

**Definition 6** (Polynomial-Scale Consecution). Let $\tau = \langle l_1, l_2, \rho_\tau, g_\tau \rangle$ be a transition and $\eta$ be an assertion map. We say that $\eta$ satisfies *polynomial-scale* consecution for $\tau$ if there exists a polynomial $q(V)$ such that

$$\rho_\tau \models (\eta(V') - q(V) \cdot \eta(V) = 0).$$

In particular, if $\deg(q) = 0$, polynomial-scale consecution reduces to constant-scale consecution.

From Definition 6, to verify whether $\eta$ satisfies polynomial-scale consecution, it suffices to check whether $\eta(V')$ can be divided by $\eta(V)$, i.e., $\eta(V) \mid \eta(V')$.

## 3. Generating Invariants of Polynomial Loops with Initial Values

In this section, we present an approach to generate polynomial equation invariants for polynomial loop programs with initial values.

Our idea is as follows.

*Step 1:* We run the loop program and get a set $S$ of sample points by recording the values of system variables at each location.

*Step 2:* We apply Buchberger-Möller algorithm to compute a Gröbner basis $\{\eta_1, \ldots, \eta_r\}$ of the vanishing ideal $\mathcal{I}(S)$ of $S$, and take the polynomials in the basis of $\mathcal{I}(S)$ (or more exactly, the corresponding polynomial equations) as candidate invariants.

*Step 3:* For each candidate invariant, we determine whether it is an invariant of the given program by checking polynomial-scale consecution.

### 3.1. Determining Divisibility between Multivariate Polynomials

We now describe Step 3, the key part of our idea in details, that is, how to verify efficiently whether a polynomial satisfies polynomial-scale consecution. Clearly, this problem is equivalent to checking divisibility between multivariate polynomials.

To fulfill this task, a straightforward way is to apply directly the multivariate polynomial division algorithm. In order to find out all polynomial invariants among the ideal basis $\eta_1, \ldots, \eta_r$, one will need to apply multivariate polynomial division algorithm $r$ times. Actually in our experiments, we find that only a few of polynomials in the ideal basis are invariants, i.e., the number of invariants is much less than $r$. For example, in Example 1 to be presented, there are six candidate invariants while only one of them is actually an invariant of the given program. Taking this special case into account, instead of directly applying multivariate polynomial division algorithm, we present a high probability algorithm to determine multivariate polynomial divisibility, combining linear transformation and univariate polynomial division. The merit of this technique lies in reducing the computational complexity.

The idea of our high probability method for determining multivariate polynomial divisibility is based on the following theorem.

**Theorem 1.** Let $f, g \in \mathcal{K}[x_1, \ldots, x_n]$ with $\deg(f) = e_1 \leq \deg(g) = e_2$, and $e_1 > 0$. Suppose that $B_2, \ldots, B_n, p_2, \ldots, p_n$ are chosen randomly and uniformly from a finite subset $W \subseteq \mathcal{K}$. Let $\tilde{f}, \tilde{g} \in \mathcal{K}[Z]$ be the univariate polynomials (in a new indeterminate $Z$) constructed from $f$ and $g$, respectively, as follows

$$\left.\begin{aligned}\tilde{f}(Z) &= f(Z, B_2 Z - p_2, \ldots, B_n Z - p_n) \\ \tilde{g}(Z) &= g(Z, B_2 Z - p_2, \ldots, B_n Z - p_n).\end{aligned}\right\} \quad (4)$$

If $f|g$, then $\tilde{f}|\tilde{g}$; or equivalently, if $\tilde{f} \nmid \tilde{g}$, then $f \nmid g$.

*Proof.* If $f|g$, then there exists a nonzero polynomial $h \in \mathcal{K}[x_1, \ldots, x_n]$ such that $g = f \cdot h$. From (4), we get $\tilde{g}(Z) = \tilde{f}(Z) \cdot \tilde{h}(Z)$ where

$$\tilde{h}(Z) = h(Z, B_2 Z - p_2, \ldots, B_n Z - p_n).$$

So we have $\tilde{f}|\tilde{g}$. □

Suppose that $\{\eta_1, \ldots, \eta_r\}$ is a set of candidate invariants. By Definition 6, to check whether $\eta_i$ satisfies polynomial-scale consecution it suffices to determine if $\eta_i(V)|\eta_i(V')$ for each $i$. Denote by $\tilde{\eta}_i(Z)$ and $\tilde{\eta}_i(Z')$ the univariate polynomials as constructed in (4) from $\eta_i(V)$ and $\eta_i(V')$, respectively. According to Theorem 1, we have the following observations:

(1) If $\tilde{\eta}_i(Z) \nmid \tilde{\eta}_i(Z')$ then $\eta_i(V) \nmid \eta_i(V')$, i.e., $\eta_i$ does not satisfy polynomial-scale consecution and then is not an invariant of the program. In such a way, the polynomials that are not invariants can be removed quickly from the invariant candidates by checking univariate polynomial divisibility.

(2) It is not necessarily true that if $\tilde{\eta}_j(Z)|\tilde{\eta}_j(Z')$ then $\eta_j(V)|\eta_j(V')$. Therefore, in the case $\tilde{\eta}_j(Z)|\tilde{\eta}_j(Z')$, to check whether $\eta_j$ is really an invariant we need to apply multivariate polynomial division to verify if $\eta_j(V)|\eta_j(V')$. However for $\tilde{\eta}_j(Z)$ and $\tilde{\eta}_j(Z')$ as constructed in (4), we will show in Theorem 2 that if $\eta_j(V) \nmid \eta_j(V')$ then $\tilde{\eta}_j(Z) \nmid \tilde{\eta}_j(Z')$ with high probability. In other words, if the multivariate polynomial $\eta_j$ does not satisfy polynomial-scale consecution, then with very low probability the corresponding univariate polynomials $\tilde{\eta}_j(Z)$ and $\tilde{\eta}_j(Z')$ satisfy $\tilde{\eta}_j(Z)|\tilde{\eta}_j(Z')$.

Before presenting the main theorem, we need several lemmas.

**Lemma 1** (Schwartz-Zippel Lemma, Schwartz (1980))**.** Let $f \in \mathcal{K}[x_1, \ldots, x_n]$ be a non-zero polynomial with $\deg(f) = e > 0$. Let $W$ be a finite subset of $\mathcal{K}$ of cardinality $|W|$ and let $r_1, r_2, \ldots, r_n$ be chosen randomly from $W$. Then we have the following probability estimate:
$$\mathrm{Prob}(f(r_1, r_2, \ldots, r_n) = 0) \leq \frac{e}{|W|}.$$

*Proof.* In the univariate case, the proof follows easily from the fact that a univariate polynomial of degree $e$ has no more than $e$ roots. The reader can refer to Schwartz (1980) for the proof in the multivariate case. □

The probability estimate in Lemma 1 will be needed to show that the multivariate polynomial and its associated univariate polynomial as constructed in (4) have the same degree with high probability.

**Lemma 2.** Let $f \in \mathcal{K}[x_1, \ldots, x_n]$ with $\deg(f) = e > 0$, $W$ be a finite subset of $\mathcal{K}$ of cardinality $|W|$, and $B_2, \ldots, B_n$ be distinct points chosen randomly from $W$. Let $\tilde{f} = f(Z, B_2 Z, \ldots, B_n Z) \in \mathcal{K}[Z]$. Then we have the following probability estimate:
$$\mathrm{Prob}(\deg \tilde{f} = e) \geq 1 - \frac{e}{|W|}.$$

*Proof.* The polynomial $f$ can be partitioned as two parts $f = f_1 + f_2$, where $f_1$ consists of the terms in $f$ with degree $= \deg(f)$ and $f_2$ consists of the terms in $f$ with degree $< \deg(f)$. If $\deg(\tilde{f}) < \deg(f)$, then we have $f_1(Z, B_2 Z, \ldots, B_n Z) = 0$. Then the probability estimate follows from Lemma 1. □

The following lemma gives an estimate of the probability that two univariate polynomials constructed as in (4) from two coprime multivariate polynomials remain coprime.

**Lemma 3.** Let $f, g \in \mathcal{K}[x_1, \ldots, x_n]$ with $\deg(f) = e_1$, $\deg(g) = e_2$ and $\gcd(f, g) = 1$. Let $W$ be a finite subset of $\mathcal{K}$ of cardinality $|W|$. Suppose that $B_2, \ldots, B_n \in \mathcal{K}$ are given such that
$$\deg(f(Z, B_2 Z, \ldots, B_n Z)) = e_1 \quad \text{and} \quad \deg(g(Z, B_2 Z, \ldots, B_n Z)) = e_2$$
where $Z$ is a new indeterminate. Let $p_2, \ldots, p_n$ be $n-1$ distinct points chosen randomly and uniformly from $W$. As in (4), let
$$\tilde{f}(Z) = f(Z, B_2 Z - p_2, \ldots, B_n Z - p_n) \text{ and } \tilde{g}(Z) = g(Z, B_2 Z - p_2, \ldots, B_n Z - p_n).$$
Then
$$\mathrm{Prob}(\gcd(\tilde{f}(Z), \tilde{g}(Z)) = 1) \geq 1 - \frac{2 e_1 e_2}{|W|}.$$

*Proof.* The proof is similar to that of Lemma 2.2 in (Kaltofen and Yang, 2007). For new variables $Z, \alpha_2, \ldots, \alpha_n$ we define the map:

$$\phi \colon \mathcal{K}[x_1, x_2, \ldots, x_n] \to \mathcal{K}[Z, \alpha_2, \ldots, \alpha_n]$$

where

$$x_1 \mapsto Z,$$
$$x_i \mapsto B_i Z - \alpha_i \quad \text{for all } 2 \leq i \leq n.$$

Namely, for any polynomial $h(x_1, x_2, \ldots, x_n)$ in $\mathcal{K}[x_1, x_2, \ldots, x_n]$,

$$\phi(h(x_1, x_2, \ldots, x_n)) = h(Z, B_2 Z - \alpha_2, \ldots, B_n Z - \alpha_n).$$

The map $\phi$ is a ring isomorphism by virtue of the inverse map

$$\phi^{-1}(Z) = x_1,$$
$$\phi^{-1}(\alpha_i) = B_i x_1 - x_i \quad \text{for all } 2 \leq i \leq n,$$

i.e., $\phi^{-1}(h(Z, \alpha_2, \ldots, \alpha_n)) = h(x_1, B_2 x_1 - x_2, \ldots, B_n x_1 - x_n)$.

Since $\phi$ is a ring isomorphism, so is $\phi^{-1}$. We can prove from $\gcd(f, g) = 1$ that $\gcd(\phi(f), \phi(g)) = 1$, i.e.

$$\gcd(\phi(f), \phi(g)) | \gcd(\phi^{-1}(\phi(f)), \phi^{-1}(\phi(g))) = \gcd(f, g) = 1.$$

Now consider the Sylvester resultant

$$\rho_1(\alpha_2, \ldots, \alpha_n) = \mathrm{Res}_Z(\phi(f), \phi(g)) \in \mathcal{K}[\alpha_2, \ldots, \alpha_n].$$

Because $\gcd(\phi(f), \phi(g)) = 1$, we have $\rho_1 \neq 0$. Remark that $\tilde{f}(Z) = f(Z, B_2 Z - p_2, \ldots, B_n Z - p_n)$ and $\tilde{g}(Z) = f(Z, B_2 Z - p_2, \ldots, B_n Z - p_n)$ are obtained by substituting $\alpha_i = p_i$ for all $2 \leq i \leq n$ into $\phi(f)$ and $\phi(g)$, respectively. From the assumption that

$$\deg(f(Z, B_2 Z, \ldots, B_n Z)) = e_1 \quad \text{and} \quad \deg(g(Z, B_2 Z, \ldots, B_n Z)) = e_2,$$

it follows that $\deg(\tilde{f}) = e_1$ and $\deg(\tilde{g}) = e_2$, and thus

$$\mathrm{Res}_Z(\tilde{f}, \tilde{g}) = \rho_1(p_2, \ldots, p_n).$$

Therefore, $\gcd(\tilde{f}, \tilde{g}) = 1$ is equivalent to $\rho_1(p_2, \ldots, p_n) \neq 0$. The probability estimate then follows from Lemma 1 and the degree estimate $\deg(\rho_1) \leq 2 \deg(\tilde{f}) \deg(\tilde{g}) = 2 e_1 e_2$. □

Now we are ready to present the main theorem.

**Theorem 2.** *Suppose that $f, g \in \mathcal{K}[x_1, \ldots, x_n]$ with $\deg(f) = e_1 \leq \deg(g) = e_2$, and that $f$ does not divide $g$, i.e., $f \nmid g$. Suppose that $B_2, \ldots, B_n, p_2, \ldots, p_n$ are chosen randomly and uniformly from a finite set $W \subseteq \mathcal{K}$. Let $\tilde{f}, \tilde{g} \in \mathcal{K}[Z]$ be the univariate polynomials constructed from $f, g$ as in (4). Then*

$$\mathrm{Prob}(\tilde{f} \nmid \tilde{g}) \geq \left(1 - \frac{2 e_1 e_2}{|W|}\right) \cdot \left(1 - \frac{e_1}{|W|}\right) \cdot \left(1 - \frac{e_2}{|W|}\right).$$

*Proof.* According to Lemma 2, the probability of picking $B_2, \ldots, B_n$ randomly from $W$ such that $\deg(\tilde{f}) = e_1$ and $\deg(\tilde{g}) = e_2$ is greater than $(1 - \frac{e_1}{|W|}) \cdot (1 - \frac{e_2}{|W|})$. Suppose that $f \nmid g$, which is equivalent to that there exists one factor $f_1$ of $f$ with $\deg f_1 > 0$ such that $\gcd(f_1, g) = 1$. Remark that $\deg \tilde{f} = \deg f$ implies that $\deg \tilde{f}_1 = \deg f_1$. Having $\gcd(f_1, g) = 1$ and under the condition $\deg \tilde{f}_1 = \deg f_1$ and $\deg \tilde{g} = \deg g$, we easily get from Lemma 3 that

$$\text{Prob}(\gcd(\tilde{f}_1, \tilde{g}) \neq 1) = 1 - \text{Prob}(\gcd(\tilde{f}_1, \tilde{g}) = 1) \leq \frac{2e_1 e_2}{|W|}.$$

If $\tilde{f}$ has a factor $\tilde{f}_1$ such that $\gcd(\tilde{f}_1, \tilde{g}) = 1$ then clearly $\tilde{f} \nmid \tilde{g}$. Therefore, the probability estimate of $\tilde{f} \nmid \tilde{g}$ is obtained as the product of two probabilities

$$\text{Prob}(\{\gcd(\tilde{f}, \tilde{g}) = 1\} \mid \{\deg \tilde{f} = e_1, \deg \tilde{g} = e_2\})$$

and $\text{Prob}(\deg \tilde{f} = e_1, \deg \tilde{g} = e_2)$. □

**Remark 3.** Stated in Theorem 2, our method is able to remove most of the non-invariant polynomials from the candidate invariants by applying univariate polynomial division, and the remaining non-invariant polynomials can be excluded by multivariate polynomial division. However, in practice, applying only univariate polynomial division can separate all the non-invariant polynomials from the actual invariants.

In the end, let us analyze the complexity of our method of determining divisibility between multivariate polynomials based on linear transformation and univariate polynomial divisibility test. The extended Euclidean algorithm is usually applied to determine divisibility between two univariate polynomials $f, g \in \mathcal{K}[x]$ with the complexity $O(\deg(f) \cdot \deg(g))$. The complexity of combining linear transformation is then given by the following theorem.

**Theorem 3.** Given $f, g \in \mathcal{K}[x_1, \ldots, x_n]$ with $\deg(f) = e_1 \leq \deg(g) = e_2$. Assume that $T = \max(T(f), T(g))$, where $T(f)$ (resp. $T(g)$) denotes the number of terms in $f$ (resp. $g$). Let $\tilde{f}$ and $\tilde{g}$ be the univariate polynomials as constructed in (4). Then the complexity of the method of determining divisibility between $f$ and $g$, based on linear transformation and the extended Euclidean algorithm is $O(T e_2 \log^2 e_2)$. In the worst case where both $f$ and $g$ are dense, the complexity is $O((n + e_2)^{e_2})$.

*Proof.* First, we analyze the cost for translating a multivariate polynomial $f$ into the univariate polynomial $\tilde{f}(Z) = f(Z, B_2 Z - p_2, \ldots, B_n Z - p_n)$ as in (4). Let $x_1^{d_1} x_2^{d_2} \cdots x_n^{d_n}$ be a term in the polynomial $f$, and denote $\bar{e} = \sum_{i=1}^{n} d_i$. The expansion of $\prod_{i=1}^{n}(B_i Z - p_i)^{d_i}$ can be computed by $O(\bar{e} \log^2 \bar{e})$ operations, according to the fan-in process method in Pan (2001). From the assumption, $T$ is an upper bound of $T(f)$ and $T(g)$, and $e_2$ is an upper bound for the degrees of all of the terms in $f$ and $g$. Therefore, the cost of computing $\tilde{f}, \tilde{g}$ is bounded by $O(T e_2 \log^2 e_2)$. If $f$ and $g$ are dense, we have $e_2 \ll T$. In this case, the cost is bounded by $O((n + e_2)^{e_2})$ since $T \leq \binom{n+e_2}{n} \leq (n + e_2)^{e_2}$, in which $\binom{n+e_2}{n}$ is the maximum number of distinct terms of a polynomial in $n$ variables and with a total degree bound $e_2$.

Clearly, the cost of applying the extended Euclidean algorithm on $\tilde{f}(Z)$ and $\tilde{g}(Z)$ is bounded by $O(e_1 e_2)$. In contrast to computing the linear transformation, this part of computation is negligible since $e_1 \leq T(f) \leq T$ in general. Therefore, the total cost of our method is $O(T e_2 \log^2 e_2)$, or is $O((n + e_2)^{e_2})$ if $f, g$ are dense polynomials. □

**Theorem 4.** (Monagan and Pearce, 2007) The complexity for computing division of multivariate polynomials using heap is $O(NM \log M)$, where $M$ is the number of terms in the quotient and $N$ is the number of terms in the divisor.

**Remark 4.** Suppose that $\eta_1, \ldots, \eta_r$ are the candidates of polynomial invariants and there are $t$ polynomials whose associated univariate polynomials satisfy $\tilde{\eta}_i(Z)|\tilde{\eta}_i(Z')$. In terms of the probability analysis shown in Theorem 2, all these $t$ polynomials are invariants of the given program with high probability. In practice, the number of invariants is usually much less than that of the candidates obtained from the vanishing ideals of sample points, i.e. $t \ll r$. Therefore, compared with $r$ times multivariate polynomial divisions, our method just needs $r$ times univariate polynomial divisions, and $t$ times multivariate polynomial divisions. Based on the complexity analysis in Theorems 3 and 4, our method of combining univariate polynomial division with partial multivariate polynomial division will be much more efficient.

*3.2. Generating Invariants of Polynomial Loops with Initial Values*

Based on the results in Section 3.1, we now present how to generate polynomial equation invariants for polynomial loops with initial values.

Let $P$ be a polynomial loop program with $n$ program variables, and $e$ an upper bound for the total degree of its potential polynomial invariants. Then we need at most $\binom{n+e}{n}$ sample points to determine a polynomial invariant in $n$ variables with total degree bound $e$. We can obtain a finite set $S$ containing no more than $\binom{n+e}{n}$ sample points by executing the program $P$.

Then, we apply Buchberger-Möller algorithm to compute the vanishing ideal

$$\mathcal{I}(S) = \langle \eta_1, \ldots, \eta_r \rangle$$

of the point set $S$, where $\eta_1, \ldots, \eta_r$ is a Gröbner basis of the ideal $\mathcal{I}(S)$. In addition, we can get the minimal degree $e'(\le e)$ of all the polynomials in $\mathcal{I}(S)$.

Consequently, we verify whether each candidate in $\{\eta_1 = 0, \ldots, \eta_r = 0\}$ is really an invariant. Clearly, all the $\eta_i$ satisfy **Initiation** condition, since they belong to the vanishing ideal of sample points of the program. Therefore, the remaining task is to determine whether $\eta_i$ satisfies **polynomial-scale consecution**, i.e., $\eta_i(V)|\eta_i(V')$. The techniques in Section 3.1 can be applied to carrying on this test efficiently.

As a result, we can either generate the polynomial invariants of total degree $\le e$ or conclude that polynomial invariants with degree $\le e'$ do not exist.

## 4. Algorithms

We now present an algorithm to generate polynomial invariants of polynomial loops with initial values, and several examples are given to illustrate the algorithm.

*4.1. Polynomial Invariant Generation of Loop Programs with Numerical Initial Values*

By omitting the guard conditions of loop programs, the following algorithm states how to generate polynomial invariants of polynomial loops with numerical initial values.

**Algorithm InvGen**
**Input:**

$P$ :    a polynomial loop program with numerical intial values

$n$ :    the number of program variables

$e$ :    an upper bound for the degree of polynomial invariants of $P$

$\sigma$ :    a term ordering on monomials of $\mathcal{K}[x_1, \ldots, x_n]$

**Step 1:** Set InvGen := NULL;
**Step 2:** Translate the loop program $P$ into a transition system $\mathrm{T} = \langle V, L, \mathcal{T}, l_0, \Theta \rangle$;
**Step 3:** Get a point set $S$ containing $\binom{n+e}{n}$ sample points by executing the loop $P$;
**Step 4:** Apply Buchberger-Möller algorithm to compute a Gröbner basis $\{\eta_1, \eta_2, \ldots, \eta_r\}$ of the vanishing ideal $\mathcal{I}(S)$ of $S$ with respect to the term ordering $\sigma$. Then

$$\{ \eta_1 = 0, \quad \ldots, \quad \eta_r = 0 \}$$

is a set of candidate invariants. Set $e' = \min\{\deg(\eta_1), \ldots, \deg(\eta_r)\}$;
**Step 5:** For $i = 1, \ldots, r$, verify whether $\eta_i = 0$ is really an invariant using Theorem 1: for $\eta_i(V)$ and $\eta_i(V')$ construct their corresponding univariate polynomials $\tilde{\eta}_i(Z)$ and $\tilde{\eta}_i(Z')$ as in (4);
  **5.1** if $\tilde{\eta}_i(Z) \nmid \tilde{\eta}_i(Z')$ then remove $\eta_i = 0$ from the candidates;
  **5.2** otherwise if $\tilde{\eta}_i(Z) | \tilde{\eta}_i(Z')$ then carry on multivariate polynomial division to check if $\eta_i(V) | \eta_i(V')$.
    **5.2.1** if $\eta_i(V) \nmid \eta_i(V')$ then remove $\eta_i = 0$ from the candidates;
    **5.2.2** otherwise if $\eta_i(V) | \eta_i(V')$ then InvGen := InvGen $\wedge \{\eta_i = 0\}$.
**Step 6:** Return the polynomial equation invariants of the program $P$.
  **6.1** If InvGen = NULL, then **return** "the polynomial equation invariants with degree $\leq e'$ do not exist."
  **6.2** Otherwise, **return** "the polynomial equation invariants of program $P$ is InvGen."

The following theorem is given to analyze the complexity of Algorithm InvGen.

**Theorem 5.** The complexity of Algorithm-InvGen is $O((n + e)^{4e+2})$, where $n$ is the number of program variables and $e$ is the upper bound for the degree of polynomial invariants.

*Proof.* Here, we ignore the cost of **Step 3**, i.e., the time for executing the loop to obtain the point set $S$. In **Step 4**, we need to apply Buchberger-Möller algorithm to compute a vanishing ideal of $\binom{n+e}{n}$ sample points, and by Remark 1, the complexity involved in **Step 4** is $O(n^2(n+e)^{4e})$. According to Theorem 3, the worst-case complexity in **Step 5**, where all the polynomials $\eta_1, \ldots, \eta_r$ are dense, is bounded by $O(r\, e\, (n+e)^{2e} \log(n+e))$. Hence, the total cost of Algorithm-InvGen is bounded by $O((n+e)^{4e+2})$. □

We give some working examples to illustrate Algorithm InvGen.

**Example 1** ((Petter, 2004; Rodríguez-Carbonell and Kapur, 2007)). Generate invariants of the following program P1:

$$(x, y) := (0, 0);$$
$$l_0: \textbf{ while } ? \textbf{ do}$$
$$(x, y) := (x + y^5, y + 1)$$
$$\textbf{end while}$$

Set $e = 7$ to be the degree bound of the polynomial invariants of P1, and let $\sigma$ be the graded lexicographical ordering $y \prec x$.
**Step 1** Set InvGen := NULL;

**Step 2** We represent the program P1 as a transition system $\langle V, L, \mathcal{T}, \{l_0\}, \Theta \rangle$:

$$\begin{cases} V = \{x, y\} \\ L = \{l_0\} \\ \mathcal{T} = \{\tau\} \text{ with } \tau = \langle l_0, l_0, x' = x + y^5 \wedge y' = y + 1, \text{true} \rangle \\ \Theta = \{x = 0 \wedge y = 0\} \end{cases}$$

**Step 3** By running the program P1 with the initial values $(0, 0)$, we get $\binom{n+e}{n} = 36$ sample points:

$$(0, 0), (0, 1), (1, 2), (33, 3), \ldots, (235306401, 34), (280741825, 35).$$

**Step 4** Apply Buchberger-Möller algorithm to get a vanishing ideal $\langle \eta_1, \ldots, \eta_6 \rangle$ of the above 36 sample points with respect to the graded lexicographical ordering $y \prec x$. We only list

$$\eta_1(x, y) = -12x + 2y^6 - 6y^5 + 5y^4 - y^2, \tag{5}$$

since the other 5 polynomials are too huge. And, the minimal degree of polynomials in the vanishing ideal is $e' = 6$;

**Step 5** Determine if $\eta_i(x, y) \mid \eta_i(x', y')$ for $i = 1, \ldots, 6$, and we get that

$$\eta_1(x, y) = -12x + 2y^6 - 6y^5 + 5y^4 - y^2 = 0$$

is an invariant of the program P1, while the other 5 candidates are not invariants of the program P1. Moreover, we can conclude that the minimal degree of the polynomial invariants $\eta(x, y) = 0$ is $e' = 6$.

Now let us consider the loop programs where both the (numerical) initial values and guard conditions are taken into account. Then the candidate invariants can be verified by solving a semi-algebraic system. Indeed, if $\eta_i(V) = 0$ satisfies the consecution condition of inductive invariants, then

$$\eta_i(V) = 0 \wedge g_\tau(V) \models \eta_i(V') = 0,$$

i.e., each real solution of $\eta_i(V) = 0 \wedge g_\tau(V)$ also satisfies $\eta_i(V') = 0$, or equivalently, the semi-algebraic system

$$(\eta_i(V) = 0) \wedge g_\tau(V) \wedge (\eta_i(V') \neq 0)$$

has no real solutions, which can be verified by Maple package `RegularChains`.

Accordingly, to generate polynomial invariants of a polynomial loop $P$ with numerical initial values and guard condition, Step 5 of Algorithm-`InveGen` needs to be revised as follows:

**Step 5'** For $i = 1, \ldots, r$, verify whether the candidate $\eta_i = 0$ is an invariant by solving the semi-algebraic system

$$(\eta_i(V) = 0) \wedge g_\tau(V) \wedge (\eta_i(V') \neq 0).$$

If the above system has no real roots, then `InvGen := InvGen` $\wedge \{\eta_i = 0\}$.

*4.2. Polynomial invariant generation of loop programs with symbolic initial values*

Algorithm `InvGen` presents how to generate polynomial invariants of loop programs with numerical initial values. However, it is quite often in the program analysis and verification that only symbolic initial values of the program variables are given. Then the boolean value of the guard conditions will rely on the (concrete) initial values of program variables. In such situations, it is hard to compute exact sample points and thus Algorithm `InvGen` cannot be applied directly.

Here, we supply two alternative ways to generate invariants for loop programs with symbolic initial values. One is to generate (symbolic) sample points by omitting the guard conditions and apply Algorithm `InvGen` with these sample points, as illustrated by the following example.

**Example 2** (Kovács (2007), Example 5.35). Generate invariants of the program P2:

$$(x, r) := (\tfrac{a}{2}, 0);$$

**while** $x > r$ **do**

$$(x, r) := (x - r, r + 1);$$

**end while**

Unlike in Example 1, the initial value of $x$ is a symbolic value, but the numerical comparison in the guard condition $x > r$ relies on the concrete value of $x$. Instead, we omit the guard condition $x > r$. A set of sample points can be collected, for example, $S = \{(\tfrac{a}{2}, 0), (\tfrac{a}{2}, 1), (\tfrac{a}{2} - 1, 2), (\tfrac{a}{2} - 3, 3), (\tfrac{a}{2} - 6, 4)\}$. Applying Algorithm `InvGen` with a degree bound $e = 2$ of the polynomial invariants, we obtain an invariant

$$2x + r^2 - r - a = 0$$

of the program P2.

An alternative approach is to combine Algorithm `InvGen` and rational function interpolation. The idea is the following.

Let $x_1, \ldots, x_n$ be the program variables of a polynomial loop $P$ and $u_1, \ldots, u_m$ the symbolic initial values of $P$. Suppose that

$$\eta = p_1(u_1, \ldots, u_m)T_1 + \cdots + p_k(u_1, \ldots, u_m)T_k = 0$$

is an invariant of the program $P$, where $T_i$ are monomials in $x_1, \ldots, x_n$ and $p_i$ are polynomials in $u_1, \ldots, u_m$. Actually, the representation of $\eta$ is not unique, for example, for any $c \in \mathcal{K} \setminus \{0\}$, $c \cdot \eta = 0$ is also an invariant. To make the representation of $\eta$ unique, we suppose that

$$\eta = T_1 + \frac{p_2}{p_1}T_2 + \cdots + \frac{p_k}{p_1}T_k.$$

The polynomial $\eta$ can be obtained using Algorithm `InvGen` and rational function interpolation method as follows. First, we assign the initial values $u_1, \ldots, u_m$ to be random values. Then for each evaluation of the symbolic initial values $u_1, \ldots, u_m$, apply Algorithm `InvGen` to obtain a polynomial invariant which involves only $x_1, \ldots, x_n$. At last, the polynomial $\eta$ can be recovered by rational function interpolation based on variable by variable interpolation and early termination techniques. More details can be found in Kaltofen and Yang (2007).

Let us look at two examples to illustrate the above method.

**Example 3** (Rodríguez-Carbonell and Kapur (2007), Example 18). Generate invariants of the program P3:

$$(x, y, u, v) := (a, b, b, a);$$
$$\textbf{while } x \neq y \textbf{ do}$$
$$\quad \textbf{if } x > y$$
$$\quad\quad (x, y, u, v) := (x - y, y, u, u + v);$$
$$\quad \textbf{else}$$
$$\quad\quad (x, y, u, v) := (x, y - x, u + v, v);$$
$$\quad \textbf{end if}$$
$$\textbf{end while}$$

In the program P3, we need to consider not only the guard condition $x \neq y$, but also two branch conditions $x > y$ and $x < y$. Let $(a, b)$ be assigned randomly, for example,

$$(a_1, b_1) = (\frac{287}{253}, \frac{751}{890}).$$

An invariant for the initial values $(a_1, b_1)$ can be generated using Algorithm InvGen:

$$\eta(x, y, u, v) = 1 - \frac{112585}{215537}xu - \frac{112585}{215537}yv = 0.$$

Assume that the invariants to be found have the form:

$$\eta_1(x, y, u, v, a, b) = 1 + \frac{p_2(a,b)}{p_1(a,b)}xu + \frac{p_3(a,b)}{p_1(a,b)}yv = 0$$

where $x, y, u, v$ are variables and $a, b$ are parameters.

Next, we recover the rational functions $\frac{p_2(a,b)}{p_1(a,b)}$ and $\frac{p_3(a,b)}{p_1(a,b)}$. By evaluating $(a, b)$ to the following numerical values, we get the corresponding invariants $\eta_i$ for $i = 2, 3, \ldots$:

$$a_2 = \tfrac{93}{122}, \quad b_2 = \tfrac{301}{992}, \quad \eta_2 = 1 - \tfrac{1952}{903}xu - \tfrac{1952}{903}yv = 0,$$

$$a_3 = \tfrac{349}{247}, \quad b_3 = \tfrac{239}{378}, \quad \eta_3 = 1 - \tfrac{46683}{83411}xu - \tfrac{46683}{83411}yv = 0,$$

$$a_4 = \tfrac{301}{6}, \quad b_4 = 3, \quad \eta_4 = 1 - \tfrac{1}{301}xu - \tfrac{1}{301}yv = 0,$$

$$a_5 = \tfrac{283}{352}, \quad b_5 = \tfrac{17}{744}, \quad \eta_5 = 1 - \tfrac{130944}{4811}xu - \tfrac{130944}{4811}yv = 0,$$

$$\vdots \qquad \vdots \qquad \vdots$$

By use of multivariate rational function interpolation based on early termination technique in (Kaltofen and Yang, 2007), an invariant of program P3 can be obtained:

$$1 - \frac{xu}{2ab} - \frac{yv}{2ab} = 0,$$

or, equivalently, $-2ab + xu + yv = 0$.

In (Rodríguez-Carbonell and Kapur, 2007), the authors obtained the same invariant as above using the fixed-point procedure, but they ignored the branch conditions $x > y$ and $x < y$. Our method takes into account the branch conditions, and are therefore more accurate in general.

**Example 4.** By modifying the algorithm (Petter, 2004) for computing sums of powers, we consider the following more complicated series of programs with symbolic initial values:

$$(x, y) := (a, b);$$

**while** *true* **do**

$$(x, y) := (x + y^k, y + 1);$$

**end while**

for $k = 1, 2, \ldots, 15$. We apply rational function interpolation method and Algorithm-`InveGen` to compute the polynomial invariants

$$\eta_k(x, y, a, b) = 0$$

that correspond to the (polynomial) assignments

$$(x, y) := (x + y^k, y + 1),$$

for $k = 1, 2, \ldots, 15$, respectively. The expressions of $\eta_k$ are given in Table 1.

## 5. Conclusions

In this paper, we present a new method for generating polynomial equation invariants for polynomial loop programs. We first generate vanishing ideals of program sample points to get candidate invariants, then provide a probabilistic method to falsify divisibility so that we can exclude quickly non-invariants in a basis of the computed vanishing ideals. Our approach avoids first-order quantifier elimination and cylindrical algebraic decomposition as well as they do not depend on any abstraction interpretation methods.

## Acknowledgment

We thank Tian Dong and Zijia Li for their helpful comments. We also appreciate the constructive comments from anonymous referees on greatly improving our paper.

## References

Buchberger, B., 1985. Gröbner-bases: An algorithmic method in polynomial ideal theory. In: Multidimensional Systems Theory - Progress, Directions and Open Problems in Multidimensional Systems. Reidel Publishing Company, Dodrecht - Boston - Lancaster, pp. 184–232.

Chen, Y., Xia, B., Yang, L., Zhan, N., 2007. Generating polynomial invariants with DISCOVERER and QEPCAD. In: Formal Methods and Hybrid Real-Time Systems. Springer-Verlag, Macao, China, pp. 67–82.

Colón, M., Sankaranarayanan, S., Sipma, H., 2003. Linear invariant generation using nonlinear constraint solving. In: Jr., W. A. H., Somenzi, F. (Eds.), CAV'2003: Computer Aided Verification. Vol. 2725 of LNCS. Springer-Verlag, Chicago, IL, USA, pp. 420–432.

Cousot, P., Cousot, R., 1977. Abstract interpretation: a unified lattice model for static analysis of programs by construction or approximation of fixpoints. In: POPL'1977: Conference Record of the Fourth Annual ACM SIGPLAN-SIGACT Symposium on Principles of Programming Languages. ACM Press, New York, NY, Los Angeles, California, pp. 238–252.

Cousot, P., Halbwachs, N., 1978. Automatic discovery of linear restraints among variables of a program. In: POPL'1978: Proceedings of the 5th ACM SIGACT-SIGPLAN Symposium on Principles of Programming Languages. ACM Press, New York, NY, Tucson, Arizona, pp. 84–96.

Floyd, R. W., 1967. Assigning meanings to programs. In: Schwartz, J. T. (Ed.), Mathematical Aspects of Computer Science. Vol. 19 of Proceedings of Symposium in Applied Mathematics. American Mathematical Socity, Providence, Rhode Island, pp. 19–32.

German, S. M., Wegbreit, B., 1975. A synthesizer of inductive assertions. IEEE Transactions on Software Engineering 1 (1), 68–75.

Kaltofen, E., Yang, Z., 2007. On exact and approximate interpolation of sparse rational functions. In: Proc. of ISSAC'07. pp. 203–210.

Kapur, D., 2004. Automatically generating loop invariants using quantifier elimination. In: ACA'04: Proc. IMACS Intl. Conf. on Applications of Computer Algebra. Beaumont, Texas.

Karr, M., 1976. Affine relationships among variables of a program. Acta Informatica 6, 133–151.

Katz, S., Manna, Z., 1976. Logical analysis of programs. Communications of ACM 19 (4), 188–206.

Kauers, M., Zimmermann, B., 2008. Computing the algebraic relations of C-finite sequences and multisequences. Journal of Symbolic Computation 43 (11), 787–803.

Kovács, L. I., 2007. Automated invariant generation by algebraic techniques for imperative program verification in Theorema. Ph.D. thesis, Research Institute for Symbolic Computation, Johannes Kepler University of Linz, Austria.

Kovács, L. I., 2008. Reasoning algebraically about P-solvable loops. In: Ramakrishnan, C. R., Rehof, J. (Eds.), TACAS'2008: Tools and Algorithms for the Construction and Analysis of Systems. Vol. 4963 of LNCS. Springer-Verlag, Budapest, Hungary, pp. 249–264.

Marinari, M. G., Möller, H. M., Mora, T., 1993. Gröbner bases of ideals defined by functionals with an application to ideals of projective points. Applicable Algebra in Engineering, Communication and Computing 4 (2), 103–145.

Möller, H. M., Buchberger, B., 1982. The construction of multivariate polynomials with preassigned zeros. In: Calmet, J. (Ed.), EUROCAM'1982: European Computer Algebra Conference. Vol. 144 of LNCS. Springer, Berlin-New York, Marseille, France, pp. 24–31.

Monagan, M. B., Pearce, R., 2007. Polynomial division using dynamic arrays, heaps, and packed exponent vectors. In: Ganzha, V. G., Mayr, E. W., Vorozhtsov, E. V. (Eds.), CASC'2007: The 10th International Workshop on Computer Algebra in Scientific Computing. Vol. 4770 of LNCS. Springer, Germany, Bonn, pp. 295–315.

Müller-Olm, M., Seidl, H., 2004a. Computing polynomial program invariants. Information Processing Letters 91 (5), 233–244.

Müller-Olm, M., Seidl, H., 2004b. Precise interprocedural analysis through linear algebra. In: POPL'2004: Proceedings of the 31st ACM SIGPLAN-SIGACT Symposium on Principles of Programming Languages. ACM Press, New York, NY, Venice, Italy, pp. 330–341.

Pan, V. Y., 2001. Structured Matrices and Polynomials: Unified Superfast Algorithms. Springer-Verlag New York, Inc., New York, NY, USA.

Petter, M., 2004. Berechnung von polynomiellen invarianten. Master's thesis, Technical University Munich, Germany.

Rebiha, R., Matringe, N., Moura, A. V., 2008. Endomorphisms for non-trivial non-linear loop invariant generation. In: Fitzgerald, J. S., Haxthausen, A. E., Yenigun, H. (Eds.), ICTAC'2008: Proceedings of the 5th International Colloquium on Theoretical Aspects of Computing. Springer-Verlag, Istanbul, Turkey, pp. 425–439.

Rodríguez-Carbonell, E., Kapur, D., 2004. Automatic generation of polynomial loop invariants: algebraic foundations. In: ISSAC'2004: Proceedings of the 2004 International Symposium on Symbolic and Algebraic Computation. ACM Press, New York, NY, Santander, Spain, pp. 266–273.

Rodríguez-Carbonell, E., Kapur, D., 2007. Generating all polynomial invariants in simple loops. Journal of Symbolic Computation 42 (4), 443–476.

Sankaranarayanan, S., Sipma, H. B., Manna, Z., 2004. Non-linear loop invariant generation using Gröbner bases. In: POPL'2004: Proceedings of the 31st ACM SIGPLAN-SIGACT Symposium on Principles of Programming Languages. ACM Press, New York, NY, Venice, Italy, pp. 318–329.

Schwartz, J. T., 1980. Fast probabilistic algorithms for verification of polynomial identities. J. ACM 27, 701–717.

Wegbreit, B., 1974. The synthesis of loop predicates. Communications of ACM 17 (2), 102–112.

Wegbreit, B., 1975. Property extraction in well-founded property set. IEEE Transactions on Software Engineering 1 (3), 270–285.

| | | |
|---|---|---|
| $\eta_1(x,y,a,b)$ | = | $y^2 - y - 2x + b - b^2 + 2a$ |
| $\eta_2(x,y,a,b)$ | = | $-6x + y - 3y^2 + 2y^3 - b + 3b^2 - 2b^3 + 6a$ |
| $\eta_3(x,y,a,b)$ | = | $4x + y^2 - 2y^3 + y^4 - b^2 + 2b^3 - b^4 + 4a$ |
| $\eta_4(x,y,a,b)$ | = | $6y^5 - 30x - y + 10y^3 - 15y^4 + b - 10b^3 + 15b^4 - 6b^5 + 30a$ |
| $\eta_5(x,y,a,b)$ | = | $-12x - y^2 + 5y^4 - 6y^5 + 2y^6 + b^2 - 5b^4 + 6b^5 - 2b^6 + 12a$ |
| $\eta_6(x,y,a,b)$ | = | $21y^5 - 42x + y + 6y^7 - 7y^3 - 21y^6 - b + 7b^3 - 21b^5$ $+21b^6 - 6b^7 + 42a$ |
| $\eta_7(x,y,a,b)$ | = | $-24x + 2y^2 - 7y^4 + 14y^6 - 12y^7 + 3y^8 - 2b^2 + 7b^4$ $-14b^6 + 12b^7 - 3b^8 + 24a$ |
| $\eta_8(x,y,a,b)$ | = | $-90x - 3y + 20y^3 - 42y^5 + 10y^9 + 60y^7 - 45y^8 + 3b$ $-20b^3 + 42b^5 - 60b^7 + 45b^8 - 10b^9 + 90a$ |
| $\eta_9(x,y,a,b)$ | = | $-3y^2 + 10y^4 - 14y^6 - 10y^9 + 2y^{10} + 15y^8 + 3b^2$ $-10b^4 + 14b^6 - 15b^8 + 10b^9 - 2b^{10} + 20a$ |
| $\eta_{10}(x,y,a,b)$ | = | $-66x + 5y - 33y^3 + 66y^5 + 55y^9 - 33y^{10} - 66y^7 + 6y^{11}$ $-5b + 33b^3 - 66b^5 + 66b^7 - 55b^9 + 33b^{10} - 6b^{11} + 66a$ |
| $\eta_{11}(x,y,a,b)$ | = | $-24x + 10y^2 - 33y^4 + 44y^6 + 22y^{10} - 33y^8 - 12y^{11} + 2y^{12}$ $-10b^2 + 33b^4 - 44b^6 + 33b^8 - 22b^{10} + 12b^{11} - 2b^{12} + 24a$ |
| $\eta_{12}(x,y,a,b)$ | = | $-2730x - 691y + 210y^{13} + 4550y^3 - 9009y^5 - 5005y^9$ $+8580y^7 + 2730y^{11} - 1365y^{12} + 691b - 4550b^3 + 9009b^5$ $-8580b^7 + 5005b^9 - 2730b^{11} + 1365b^{12} - 210b^{13} + 2730a$ |
| $\eta_{13}(x,y,a,b)$ | = | $-420x - 210y^{13} - 691y^2 + 2275y^4 + 30y^{14} - 3003y^6$ $-1001y^{10} + 2145y^8 + 455y^{12} + 691b^2 - 2275b^4 + 3003b^6$ $-2145b^8 + 1001b^{10} - 455b^{12} + 210b^{13} - 30b^{14} + 420a$ |
| $\eta_{14}(x,y,a,b)$ | = | $-90x + 105y + 105y^{13} - 691y^3 + 6y^{15} + 1365y^5 + 715y^9$ $-45y^{14} - 1287y^7 - 273y^{11} - 105b + 691b^3 - 1365b^5 + 1287b^7$ $-715b^9 + 273b^{11} - 105b^{13} + 45b^{14} - 6b^{15} + 90a$ |
| $\eta_{15}(x,y,a,b)$ | = | $-48x + 420y^2 - 24y^{15} - 1382y^4 + 60y^{14} + 1820y^6 + 572y^{10}$ $-1287y^8 + 3y^{16} - 182y^{12} - 420b^2 + 1382b^4 - 1820b^6 + 1287b^8$ $-572b^{10} + 182b^{12} - 60b^{14} + 24b^{15} - 3b^{16} + 48a$ |

Table 1: The Invariants in Example 4